\documentclass[manuscript,screen]{acmart}

\AtBeginDocument{%
  }

\setcopyright{acmlicensed}
\copyrightyear{2018}
\acmYear{2018}
\acmDOI{XXXXXXX.XXXXXXX}

\acmConference[Conference acronym 'XX]{Make sure to enter the correct
  conference title from your rights confirmation emai}{June 03--05,
  2018}{Woodstock, NY}
\acmISBN{978-1-4503-XXXX-X/18/06}

\usepackage{makecell}
\usepackage{tabularx}
\usepackage{tabulary}
\usepackage[]{mdframed}
\usepackage{longtable}
\usepackage{tabularray}
\usepackage{rotating}

\begin{document}

\title{Challenges in GenAI and Authentication: a scoping review}

\author{Wesley dos Reis Bezerra}
\authornote{All authors contributed equally to this research.}
\email{wesleybez@gmail.com}
\orcid{0000-0002-6098-7172}

\author{Laís Machado Bezerra}
\authornotemark[2]
\email{lais.machado1@gmail.com}
\orcid{0000-0003-3331-7277}

\author{Carlos Becker Westphall}
\authornotemark[1]
\email{carlosbwestphall@gmail.com}
\orcid{0000-0002-5391-7942}

\affiliation{%
  \institution{PPGCC/UFSC\authornotemark[1]}
  \city{Florianópolis}
  \state{Santa Catarina}
  \country{Brazil}
}
\affiliation{%
  \institution{SENAI/SC\authornotemark[2]}
  \state{Santa Catarina}
  \country{Brazil}
}

\renewcommand{\shortauthors}{WR Bezerra, LM Bezerra, CB Westphal}

\begin{abstract}
Authentication and authenticity have been a security challenge since the beginning of information sharing, especially in the context of digital information. With the advancement of generative artificial intelligence, these challenges have evolved, demanding a more up-to-date analysis of their impacts on society and system security. This work presents a scoping review that analyzed 88 documents from the IEEExplorer, Scopus, and ACM databases, promoting an analysis of the resulting portfolio through six guiding questions focusing on the most relevant work, challenges, attack surfaces, threats, proposed solutions, and gaps. Finally, the portfolio articles are analyzed through this guiding research lens and also receive individualized analysis. The results consistently outline the challenges, gaps, and threats related to images, text, audio, and video, thereby supporting new research in the areas of authentication and generative artificial intelligence.    
  
\end{abstract}


\keywords{Do, Not, Us, This, Code, Put, the, Correct, Terms, for,
  Your, Paper}

\received{20 February 2007}
\received[revised]{12 March 2009}
\received[accepted]{5 June 2009}

\maketitle

\section{Introduction}
\label{S:1}

Verifying the authenticity of messages and their true origin is currently a challenge for computer users. This applies to users of cell phones, computers, wearable devices, sensors, and other Internet of Things (IoT) devices. Furthermore, this has become even more challenging due to the advancements brought about by access to Generative Artificial Intelligence (GenAI) in recent years, which has been achieved through various tools and techniques.

With access to GenAI tools, some malicious users have been carrying out scams using this technology, such as impersonation. This is a way of impersonating other people by adopting their characteristics, behavior, or demeanor. This form of security breach is especially complex to address in situations where users have little knowledge or familiarity with GenAI tools, for example, older users with limited computer skills. The main problems for end users include a large number of scams that use AI-generated images (\cite{hu2024transfer}), videos (\cite{ghildyal2024quality}), and audio (\cite{kurshan2024ai}) to confuse people close to the impersonated person (\textit{impersonated}) and lead them to make errors in judgment that often lead to financial losses (\cite{neupane2023impacts}).

Currently, not all companies are prepared for this type of situation and security breach. Some companies still need to evolve considerably in basic security issues, and planning how to address breaches originating from Artificial Intelligence (AI) is not a trivial issue to be overlooked. As a further aggravating factor, because it is a very recent topic, our legislation (Brazil) is not fully adapted to this new type of threat. This means that both awareness-raising efforts and punitive measures in the event of a material breach lack tangible and easily visible effects on society. Therefore, a more in-depth analysis of the potential security impacts of generative AI tools is necessary. This analysis should be based on the state-of-the-art in authentication and propose methods for effectively identifying breach attempts using generative AI.

This work aims to discuss various aspects involving the authentication and authenticity of artifacts produced by generative AI. This discussion is conducted through a scoping review that presents a portfolio representing the state-of-the-art in authentication/authenticity and GenAI, along with their challenges. The main contributions include:

\begin{itemize}
\item a survey of the state of the art through a \textit{scoping review} literature review;
\item an outline and discussion of the main issues associated with GenAI and authentication; 
\item an analysis of the portfolio resulting from the literature review;
\end{itemize}

Through these contributions, we will support the development of programs that are better equipped to address the new challenges of Generative AI (GenAI). A greater understanding of this emerging area will enable end-users and researchers to be better prepared to face future challenges. It will also facilitate a better understanding of the attack possibilities (attack surface) of the technologies involved in development within this new scenario.

The remainder of the work is organized as follows: Section \ref{sec_methodology} presents the methodology for developing the work, which outlines the data acquisition procedures. This is followed by a synthesis of the data and articles collected in Section \ref{sec_data_systhesis}. In Section \ref{sec_analysis}, guiding questions are used to analyze the articles belonging to the portfolio. Finally, Section \ref{sec_conclusion} presents the conclusions and future work.

\section{Methodology}
\label{sec_methodology}

This section outlines the methodology employed in this work. It discusses the checklist used to develop the Scoping Review protocol, the framework used to report this review, the location of the files used to track the workflow, the tools used to manage the work, and other relevant details. In general, this section presents the entire framework of methods, frameworks, and tools used to develop the work.

This work adopted the Preferred Reporting Items for Systematic Review and Meta-Analyses extension for Scoping Review (PRISMA-ScR) \cite{tricco2018prisma} for its development. This scoping review checklist is an extension of the PRISMA \cite{moher2009preferred} specific to scoping reviews. Its purpose is to bring more formalization and reduce bias in research using it. It consists of 20 mandatory items and two optional items and has gained notoriety based on the work of McGowan et al. \cite{mcgowan2020reporting}.

The Joanna Briggs Institute methodology framework \cite{peters2015joanna, peters2015guidance}, which presents a well-established guide for developing scoping reviews, was also used. This type of review has a less rigid structure than the systematic literature review (SLR) and is less rigorous regarding the documents used as a basis, and can also include gray literature, manuals, and patents.

Another important aspect to be mentioned in this methodology is the culture of open science \cite{hosseini2025open,mirowski2018future,pordes2007open}. With the same worldview of open source \cite{hoffmann2024value, fitzgerald2006transformation} and open hardware design (open hardware) \cite{bkedkowski2024open,powell2012democratizing}, open science brings visibility and transparency to the process of doing science and relies on several tools to support it. One of these tools is the Open Science Framework (OSF) \cite{foster2017open}, which provides a platform for publicizing scientific studies and enabling better analysis of the quality of the process, thus reducing the risk of bias. 

The Rayyan web tool \cite{ouzzani2016rayyan} was chosen as the management tool for the set of articles used. This tool facilitates collaborative work among research team members and generates graphs that illustrate the selection process.

\subsection{Objectives and Research Question}
\label{S:3}

This work focused on security aspects of GenAI, providing a broader view of the challenges and how they materialize in the real world. Some authentication and authenticity challenges have currently emerged in the GenAI field; however, it is important to have a general overview of the applicability of these challenges to existing cyber-physical systems. This general overview allows for a plan for future work to address the challenges encountered.

From this perspective, a broad research question was developed that allows for a focus on applications and practical aspects of GenAI in breaching authentication/authenticity security in cyber-physical systems. This question can be seen below.

\centerline{
\begin{minipage}{.75\textwidth}
\hfill \break
\hfill \break
\textit{"What are the security challenges for authentication brought about by the advent of Generative Artificial Intelligence and their impact on threat generation that can be found in the most relevant work today?" }
\end{minipage}
}
\hfill \break

The hypothesis of creating a document that would support future research in the area was the starting point, and therefore, a more comprehensive question was formulated. The hope is that this document will support the development of new technologies that prepare cyber-physical systems for the challenges posed by advances in Generative Artificial Intelligence. However, to arrive at an adequate answer to this question, we must segment the purposes of this research and seek support from the six guiding questions listed below.

\begin{itemize}
    \item[\textbf{Q1}] What are the most relevant works in the area of authentication and GenAI?
    \item[\textbf{Q2}] What are the security challenges encountered in these works?
    \item[\textbf{Q3}] What are the attack surfaces most targeted by the attacks?
    \item[\textbf{Q4}] What are the main threats listed in the works found?
    \item[\textbf{Q5}] What are the main proposed solutions?
    \item[\textbf{Q6}] What are the gaps found in the selected studies? 
\end{itemize}

Question \textbf{Q1} provides an overview of the most relevant research work within the research area, along with an outlook on advances, challenges, and future work to be explored within the framework of these documents. \textbf{Q2} focuses specifically on security problems or challenges without limiting areas of activity or the type of security addressed. \textbf{Q3} and \textbf{Q4} seek to understand which types of systems or resources are the main target of adversaries and the approach to exploiting vulnerabilities in them, respectively. Furthermore, question \textbf{Q5} provides an overview of existing solutions within their specific areas and their characteristics. Finally, \textbf{Q6} outlines an overview of existing research opportunities within the area and the resolution of still unresolved challenges.

\subsection{Query string and Databases}
\label{S:4}

The proposed query string (\ref{eq:query_string}) provides a broader view of the field and is based on only three terms. The first term is Generative Artificial Intelligence ("genai"), which has a broader focus and does not restrict the type of use or result produced by the AI in question. The second term is "authentication," which serves both to identify the authentication of cyber-physical systems and the authentication of artifacts (proof of their authenticity). Finally, the term "challenges" presents us with a way to both present existing solutions to the challenges and to search for documents that present challenges that are still being researched. Therefore, we chose to use the logical operator AND between the terms mentioned to bring together these broader areas and delimit our interest within them.

\begin{equation}
genAI\; AND\; authentication\; AND\; Challenges
\label{eq:query_string}
\end{equation}

\hfill \break

To select the databases, we chose three databases that together represent a large portion of the most well-known conferences and journals in the field of computing. The selected databases were IEEExplorer, Scopus, and the ACM Digital Library, all of which allow the use of a specific search language and provide tools for advanced searches and filters. Furthermore, these databases allow the export of search content in a standard format, such as BibTeX or CSV.

\begin{table}[h]
\centering
\begin{tabular}{lll}
\toprule
\textbf{Database} & \textbf{Quantity} & \textbf{Percentage}\\
\midrule
IEEExplorer & 8 documents & 9.5\%\\
Scopus & 15 documents & 17.6\%\\
ACM & 62 documents & 72.9\%\\
\midrule
Total & 85 documents & 100\%\\
\bottomrule
\end{tabular}
\caption{Articles by database}
\label{tab:database_search}
\end{table}

As shown in Table \ref{tab:database_search}, a total of 85 documents were retrieved through the search. For the IEEExplorer database, a total of eight documents (9.5\%) were retrieved, while for Scopus, 15 documents (17.6\%) were obtained, and ACM had the largest number of documents, with 62 documents (72.9\%). Although the sample size is not large, the documents obtained from the three databases represent a concise and non-segmented set of articles.

\subsection{Inclusion and Exclusion Criteria}
\label{S:5}

From this set of documents, criteria were selected for inclusion in the study. Inclusion criteria (Table \ref{tab:inclusion_criteria}) comprised a total of four criteria, and exclusion criteria (Table \ref{tab:exclusion_criteria}) comprised six items for all selected articles. These criteria aim to establish a well-formalized selection process that is easy to understand for all those involved in the "document selection" process. These criteria are further detailed below.

\begin{table}[h]
\centering
\begin{tabular}[t]{ll}
\toprule
\# & \textbf{Inclusion}\\
\midrule
\textbf{I1} & \makecell[l]{The article must be directly related to the use of GenAI that allows the exploitation of an authentication flaw}\\
\textbf{I2}& \makecell[l]{Papers that demonstrate the attack surface, attack techniques, and report cases of exploited or potentially \\exploitable flaws will be included.}\\
\textbf{I3}& \makecell[l]{Papers with a five-year time horizon (prior to the date of this study),\\ that is, from 2020 to 2024.}\\
\textbf{I4}& \makecell[l]{Papers written in English only will be selected.}\\
\bottomrule
\end{tabular}
\caption{Inclusion Criteria}
\label{tab:inclusion_criteria}
\end{table}

The first \textbf{inclusion criterion} is \textbf{I1}, which selects documents specifically related to authentication, as false positives may appear during the search. This is followed by \textbf{I2}, which includes the set of documents that present specifications on \textit{attack surfaces}, attacks, and cases. The third criterion, \textbf{I3}, addresses the temporal aspect, selecting a four-year time window (2020 to 2024). The fourth criterion, \textbf{I4}, focuses on the content distribution method in terms of publication language and limits the set to only English-language documents.

\begin{table}[h]
\centering
\begin{tabularx}{\textwidth}[t]{ll}
\toprule
\# & \textbf{Exclusion}\\
\midrule
\textbf{E1} & \makecell[l]{Short articles, abstracts, and other works that do not address a complete study of the topic will be excluded.}\\
\textbf{E2} & \makecell[l]{Works that summarize, summarize, or superficially address the topic will be excluded.}\\
\textbf{E3} & \makecell[l]{Works that are not scientific, such as patents, white papers,\\ among others, will be excluded.}\\
\textbf{E4} & \makecell[l]{Replicated works or those with little difference in their text but that have been published in more than \\one publication will be excluded.}\\
\textbf{E5} & \makecell[l]{Works that are not in the field of computer science or a related field will be excluded.}\\
\textbf{E6} & \makecell[l]{Works published before 2020 will not be included.}\\
\bottomrule
\end{tabularx}
\caption{Exclusion Criteria}
\label{tab:exclusion_criteria}
\end{table}

The \textbf{exclusion criteria} are used to support the removal of a document from the final group of articles, called a portfolio. Starting with the in-depth studies \textbf{E1}, \textbf{E2}, and \textbf{E3}, which initially yielded no significant results and were reported in summary or brief form, and works lacking scientific scope were excluded, respectively. Regarding the \textbf{E4} multi-media report of the same study, only the most comprehensive document was selected, and the others were discarded. Furthermore, \textbf{E5} addresses works that were not proposed within the scope of computer science or a related field and were not included in the final selection. Finally, regarding the temporality criterion (\textbf{E6}), works prior to 2024 will not be included in the final portfolio.

Regarding the risk of bias, this risk was mitigated due to the use of a well-recognized method of proven quality. The use of a data source where the articles were subjected to peer review also reinforces the mitigation of bias in the research.

\section{Data Sysnthesis}
\label{sec_data_systhesis}

\begin{figure}
\centering
\includegraphics[width=0.75\linewidth]{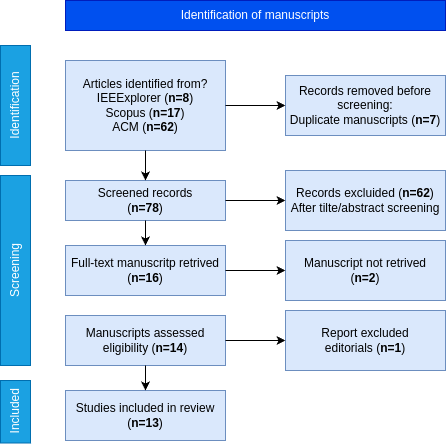}
\caption{PRISMA Workflow}
\label{fig:prisma_workflow}
\end{figure}

An overview of the data analysis and synthesis is presented in Figure \ref{fig:prisma_workflow}, which illustrates the PRISMA Workflow for this study. The extracted data followed a set of fields described in the Appendix \ref{sc:data_extraction}.

Data selection was performed through the evaluation of 85 papers selected from three databases and was supported by the Rayyan tool \footnote{\url{https://rayyan.ai/}}. Initially, duplicate papers were extracted; that is, if any of the 14 papers found were selected, none presented more than one copy in the adopted selection. From this point on, title and abstract screening were performed, resulting in the removal of a total of 62 documents. Of the 16 remaining articles, copies of 14 of them were obtained. After a thorough reading and evaluation, one document was excluded because it was an editorial and not a scientific publication. In conclusion, a portfolio of 13 articles was obtained for data extraction and analysis.

Twelve fields were chosen as data extraction fields, as shown in Table \ref{tab:field_description} in Appendix \ref{sc:data_extraction}. A sequential identifier was created for identification purposes, and the authors' names and publication title were maintained. For methodology, the following were selected: the problem addressed, the object of study, the application context, the study type, the area, the topic, and the approach used. The database and number of participants were selected as resources. Further details can be found in the Appendix (\ref{sc:data_extraction}), which contains a tabulation of the extracted data.

\subsection{Analysis of Guiding Questions}
\label{subsec_questions}

The final portfolio will be analyzed in accordance with the guiding questions outlined in this study, which are listed in the Methodology section (Section \ref{sec_methodology}).

\subsubsection{Q1. What are the most relevant works in the area of authentication and GenAI?}

Starting with \textbf{Q1}, we can affirm that the selected works are the most relevant within the set of works available at the time of the search. However, because this is an area undergoing significant research expansion, the dynamics among researchers, authors, co-authors, and funders tend to change. An overview of the most important works is presented in Table \ref{tab:portfolio_summary}, which summarizes our portfolio data.

\begin{table}[h] 
\centering 
\begin{tabularx}{\textwidth}{@{}llX@{}} 
\toprule
\textbf{\#} &\textbf{Ref.} &\textbf{Article}\\
\midrule
1 & \cite{datta2021perfect} &A Perfect Storm: Social Media News, Psychological Biases, and AI\\
2 &\cite{lewicki2023out} &Out of Context: Investigating the Bias and Fairness Concerns of “Artificial Intelligence as a Service”\\
3 &\cite{yu2023antifake} &AntiFake: Using Adversarial Audio to Prevent Unauthorized Speech Synthesis\\
4 &\cite{hutiri2024not} &Not My Voice! A Taxonomy of Ethical and Safety Harms of Speech Generators\\
5 &\cite{capasso2024comprehensive}&A Comprehensive Survey on Methods for Image Integrity\\
6 &\cite{liu2024provcam} &ProvCam: A Camera Module with Self-Contained TCB for Producing Verifiable Videos\\
7 &\cite{korobenko2024towards} &Towards a Privacy and Security-Aware Framework for Ethical AI: Guiding the Development and Assessment of AI Systems\\
8 &\cite{ristovska2023ways}&Ways of seeing: The power and limitation of video evidence across law and policy\\
9 &\cite{han2024uncovering} &Uncovering Human Traits in Determining Real and Spoofed Audio: Insights from Blind and Sighted Individuals\\
10 &\cite{punnappurath2024advocating}&Advocating Pixel-Level Authentication of Camera-Captured Images\\
11 &\cite{bird2024cifake}&CIFAKE: Image Classification and Explainable Identification of AI-Generated Synthetic Images\\
12 &\cite{bianco2023identifying}&Identifying AI-Generated Art with Deep Learning\\
13 &\cite{shoaib2023deepfakes}&Deepfakes, Misinformation, and Disinformation in the Era of Frontier AI, Generative AI, and Large AI Models\\
\bottomrule 
\end{tabularx} 
\caption{Portfolio - most representative articles} 
\label{tab:portfolio_summary}
\end{table}

\subsubsection{Q2. What security challenges have you encountered in these projects?}

Regarding question \textbf{Q2}, which addresses the challenges encountered, we can categorize them into: \textit{deepfake}, lack of regulation, bias and lack of \textit{fairness}, information bias anchoring (IBA), fake news, swatting attacks, other risks involving false emergency alarms, among others. As shown, there are many challenges, and their solutions depend on different approaches. For a better understanding, the portfolio projects and their specific challenges are listed below:

\begin{itemize}

    \item\cite{datta2021perfect} highlights the main challenges of combating information bias, which is related to both technological aspects that hinder the identification of fake news and psychological aspects that make users inclined to believe fake news.

    \item\cite{lewicki2023out} For the authors, the challenges are linked to AI regulation, mainly regarding aspects such as data protection and non-discrimination.

    \item\cite{yu2023antifake} presents the main challenge of preventing the use of audio files published by users to generate deepfakes.

    \item\cite{hutiri2024not} presents the stress caused to citizens and the unavailability of services due to the use of deepfakes audio in phone calls.

    \item\cite{capasso2024comprehensive} presents problems arising from image forgery, which pose challenges in its identification. Since forgery identification solutions are specific to each scenario, there is currently no general solution to the problem. 
    
    \item\cite{liu2024provcam} highlights the problem of ensuring video authenticity even with technological advances in GenAI image generation. As models have advanced, passively detecting altered or AI-generated videos has become increasingly difficult.

    \item\cite{korobenko2024towards} presents privacy as one of the main challenges and details this issue in four dimensions: data, technology, people, and processes.

    \item\cite{ristovska2023ways} explores the use of videos as evidence by police and law enforcement in the era of digital media synthesis.

    \item\cite{han2024uncovering} identifies verification of audio authenticity based on voice characteristics. Both sighted and blind people evaluated these.

    \item\cite{punnappurath2024advocating} raises the challenge of identifying hallucinations generated by automatic AI correction modules currently found in photographic devices.

    \item for \cite{bird2024cifake}, the main problem is the inability to identify fakes in photographs, that is, the inability to easily identify a computer-generated image.

    \item\cite{shoaib2023deepfakes} addresses the identification of \textit{fakes} generated by Large Language Models (LLM) used for manipulation and other types of violations on the internet.

    \item for the authors of \cite{bianco2023identifying} the main challenges are the identification of human-made artworks and artworks made by GenAI.
    
\end{itemize}

\subsubsection{Q3. What are the attack surfaces most targeted by these attacks?}

Most attacks occur on the web, specifically on social media. The spread of deepfakes and fake news is a growing problem and has become increasingly difficult to identify for users without technical expertise. Due to the increased image, video, and audio generation capabilities provided by AI tools, many people have difficulty identifying whether AI produced a piece of media or not. Below is the analysis for each publication:

\begin{itemize}

    \item \cite{datta2021perfect} presents social media tools as a platform for the dissemination of fake news and the anchoring of bias.

    \item \cite{lewicki2023out} For the authors, the threats present themselves as more general challenges to fairness. We can say that the topic addresses three areas: the provider, the user, and legislation.

    \item \cite{yu2023antifake} presents surfaces that use voice as information, such as voice authenticators, phone calls, audio messages, and public audio publications.

    \item \cite{hutiri2024not} attacks are carried out through phone calls and occur mainly in schools and citizen service centers.

    \item \cite{capasso2024comprehensive} the use of generated images can affect communication between people, the safety of children and adolescents, system authentication, political systems, among others.

    \item \cite{liu2024provcam} For the authors, the dissemination of altered or AI-generated videos on social media has been a problem. This problem is further exacerbated by the complexity of identifying the alteration of these videos.

    \item for \cite{korobenko2024towards}, according to the framework proposal, computational resources, data, and system (technology) are the chosen attack surfaces.

    \item for \cite{ristovska2023ways}, the legal system and the police are directly impacted, since authenticity challenges directly impact these systems.

    \item \cite{han2024uncovering} This study does not directly focus on attacks but comments on the vulnerability of people who must base their identification of emotions and individuals solely on voice: blind people.

    \item \cite{punnappurath2024advocating}'s work highlights the authenticity challenges faced by modern image acquisition devices and AI post-processing.

    \item \cite{bird2024cifake} primarily focuses on image-based media, such as art, photography, and any other medium that disseminates these types of digital artifacts.

    \item \cite{shoaib2023deepfakes} highlights information publishing platforms as one of the most affected, especially social media.

    \item \cite{bianco2023identifying}'s most affected entities are companies that price, evaluate, or trade works that humans may not have created.

\end{itemize}

\subsubsection{Q4. What are the main threats listed in the papers found?}

Although there is a large set of relevant threats when it comes to security and Generative AI, here are the main threats existing in our portfolio and selected by the authors. They will be presented and briefly commented on in their specific papers, listed below:

\begin{itemize}

    \item \cite{datta2021perfect} highlights the spread of fake news, misinterpretation, the uncertainty of significant, frequent, and anonymous information sets, and internal bias.

    \item \cite{lewicki2023out} presents an argument that can be summarized as the lack of legislation and monitoring to ensure accountability for the lack of fairness in the use of AI as a service.

    \item \cite{yu2023antifake} discusses the problems posed by the use of deepfake for voice synthesis. Threats include conducting financial scams, compromising security and privacy, spreading hate speech and disinformation, and violating voice authentication.

    \item \cite{hutiri2024not} presents risks such as swatting attacks, which use synthesized voices to spread false alarms or call emergency services (e.g., fire departments), causing some service units to be unavailable due to their use in false alarms.

    \item \cite{capasso2024comprehensive} highlights some of the main threats, including encouraging child prostitution, political misconduct, falsifying medical images, and violating system authentication.

    \item \cite{liu2024provcam}, this paper presents image damage and misinformation as the main problems arising from the difficulty in identifying the generation and alteration of videos made by GenAI.

    \item \cite{korobenko2024towards} lacks guarantees of the process's operation, access to sensitive information, and transparency in the process, leading to privacy violations. In general, these are problems related to privacy.

    \item \cite{ristovska2023ways} the lack of safeguards for this type of situation, the compromise of evidence, and the lack of preparation of legal systems to address this type of situation.

    \item \cite{han2024uncovering} the paper discusses the risks of impersonation, audio replay, speech synthesis, and voice conversion. These situations are analyzed by individuals in a study presented in the paper.

    \item for \cite{punnappurath2024advocating} altering the information contained in the image can lead to misidentification of some elements of the image. This can lead to decision-making based on information inserted by an AI module's hallucination.

    \item \cite{bird2024cifake} the main threat lies in the failure to identify a computer-generated image. From this point on, several ethical, professional, and legal issues arise in the use of this type of technology.

    \item \cite{shoaib2023deepfakes} cites the effects on democracy and public opinion, impacts on privacy and personal security, consequences for media and journalism, erosion of public trust, and ethical and legal dilemmas as threats.

    \item \cite{bianco2023identifying} incorrect identification can lead to problems in assessing the value of a work, for example. Furthermore, credibility issues can arise in cases where works are misjudged.

\end{itemize}

\subsubsection{Q5. What are the main proposed solutions?}

Each problem demands a different specific solution approach. However, in general, any solution must involve a technological, legal, or cultural approach. The technological approach is critical in identifying digitally generated artifacts, but only education and legal penalties can resolve the situation in the long term. More informed people, with a more open outlook on life and receptive to different opinions, are less susceptible to being deceived and to replicating fake news. In addition, there is a need for legal accountability for the generation and dissemination of fakes by the participating parties. A more detailed discussion per portfolio article follows.

\begin{itemize}

    \item \cite{datta2021perfect} brings transparency and the reduction of internal bias as objectives for solutions. As a tangible technological solution, it brings the use of eXplainable Artificial Intelligence (XAI) in generative AI.

    \item \cite{lewicki2023out} presents regulation, surveillance, transparency, and accountability as part of the solution to addressing fairness issues in AI platforms.

    \item \cite{yu2023antifake} proposes using the solution presented in the article, which processes audio before it is made public. This processing aims to embed information that prevents deepfakes from being overly precise and difficult to identify.

    \item \cite{hutiri2024not} presents a framework that allows for a better understanding of the data these threats can generate and follows this classification into three dimensions: data types, specific data, and motivation for the harm.

    \item \cite{capasso2024comprehensive} presents two types of forgery detection: active detection and passive detection. Active detection can be used through watermarking, digital signatures, and encryption techniques. Passive detection, on the other hand, consists of applying detection techniques through feature extraction to identify image tampering, identify nonconformities, and distinguish between natural and computer-generated images.

    \item \cite{liu2024provcam} proposes a solution based on secure hardware that utilizes cryptographic techniques to ensure the origin of video generation and the integrity of the image sequence that composes it.

    \item \cite{korobenko2024towards} the creation of a framework for the ethical use of AI with security and privacy. This can be expressed through the recommendations proposed by the authors.

    \item \cite{ristovska2023ways} improvements in countries' legal systems and the adoption of cutting-edge technologies that can address these types of situations where video authenticity issues may arise.

    \item \cite{han2024uncovering} proposes the possibility of guaranteeing the origin of audio files and the use of digital watermarks that are perceptible to humans.

    \item \cite{punnappurath2024advocating} demonstrates a possible solution through the use of spatial masking for altered pixels, the use of metadata, and the pixel-by-pixel identification of the changes made.

    \item \cite{bird2024cifake} proposes the solution by using its image dataset in conjunction with XAI and a proposed method for identifying fakes.

    \item \cite{shoaib2023deepfakes} presents a framework that divides responsibilities and promotes collaboration between the parties involved, incorporating technology and legal aspects to solve this problem.

    \item \cite{bianco2023identifying} the authors propose the use of a dataset proposed by them, and the adoption of XAI and Gradient Class Activation Mapping (Grad-CAM) in the identification of AI-generated art.
    
\end{itemize}

\subsubsection{Q6. What gaps were found by the selected studies?}

Due to the recent adoption of GenAI by the general public, new forms of security breaches are emerging every day. Added to this are advances in models and the quality of generated artifacts, which are becoming increasingly closer to reality. In this scenario, many research opportunities are needed to address the gaps that still exist in the literature. Below, we will list some gaps found in each selected article.

\begin{itemize}

    \item \cite{datta2021perfect} proposes the use of XAI as part of the content generation solution. This would be a way to identify whether the generated content is sufficiently transparent and free of bias.

    \item \cite{lewicki2023out} presents the lack of regulation and transparency as important gaps in the topic of fairness.

    \item \cite{yu2023antifake} applying pre-processing to audio before publishing remains complex for the end user. The lack of different samples of adequate length also poses challenges for research, as well as ethical issues and bias in approaching the topic.

    \item \cite{hutiri2024not} highlights the need for: distinguishing between testable and untestable data, studying model outputs, and understanding the potential harm caused by exposure, ways to mitigate these types of harm, and legal support for accountability through policies and civil society interventions.

    For the authors of \cite{capasso2024comprehensive}, research opportunities exist primarily in the new areas of image modification detection, namely image anomaly detection and deep anomaly detection. The authors also classify the use of deep learning in anomaly detection as one of the main trends.

    \item in \cite{liu2024provcam}, the authors outline improvement proposals for the application of their solution to real devices such as drones, robots, cameras, cell phones, among others.

    \item also for \cite{korobenko2024towards}, the need to evolve the framework in line with the evolution of algorithms and the application of AI in various areas.

    \item in \cite{ristovska2023ways}, there are opportunities for advancement in countries' legal systems, in the development of new technologies for identifying deepfakes and authenticating videos to be used as evidence.

    \item for \cite{han2024uncovering}, the advancement of audio synthesis has generated results very close to reality, allowing even individuals more sensitive to the different characteristics of voice to be deceived. Research into source identification and human-identifiable watermarks is necessary.

    \item \cite{punnappurath2024advocating} highlights research opportunities in how to ensure that the mask and metadata generated during acquisition are not altered, as well as how to store such information.

    \item \cite{bird2024cifake} argues how gaps represent opportunities for using different classification techniques within their dataset, the use of other XAI approaches, and the continued evolution of the dataset to keep pace with new advances in image generation.

    \item \cite{shoaib2023deepfakes} proposes continued improvement in identification methods to keep pace with GenAI developments, adaptability to address new forms of disinformation, adoption and implementation of standards, and education of information consumers. The work also complements the list of six open research opportunities: technological advancements, global cooperation, public engagement, behavioral insights, economic models, and technological innovations.

    \item already \cite{bianco2023identifying} proposes an ensemble approach that combines multiple models for a more robust output.
    
\end{itemize}

\section{Portfolio Analysis}
\label{sec_analysis}

The work of \textbf{Datta et al}. \cite{datta2021perfect} raises questions about how AI can exacerbate bias in social media. This work focuses on \textit{fake news} and text-based artifacts. This work introduces the concept of IBA and promotes an approach through a technological lens combined with behavioral psychology.

According to the authors, aspects of generating content quickly, diversely, and anonymously can generate information overload, leading to uncertainty and difficulty in identifying fake news for users. Added to this are psychological aspects that lead users to be selective when choosing information sources and to misinterpret information based on their pre-existing biases.

Several approaches are listed as solutions. Starting with eXplainable Artificial Intelligence to allow the user to understand the AI's inference process. Adopting a multidisciplinary process facilitates analysis from more than one perspective. Increasing transparency and reducing uncertainty would allow for fewer problems interpreting the facts presented. Retraining the AI to reduce bias, even if more expensive, would help with fake news. Finally, reducing intrinsic bias is a complex problem, as it addresses values learned in childhood, relationships with parents, and even linguistic aspects.

The next work in the portfolio is by \textbf{Lewicki et al.} \cite{lewicki2023out}, which presents bias and fairness in AI as a Service provisioning. In this type of service, the company can hire an AI provider to meet its processing needs. In its taxonomy, the paper presents three service provision models: autoML platforms, AI APIs, and fully managed AI services.

Furthermore, the need for fairness in algorithms and AI service provision is essential, since users may not provide training data but instead use a previously trained knowledge base. The paper discusses what fairness is and how considering it a contextual concept can create problems for this type of service provision. For the authors, each area presents a challenge: autoML platforms face the challenge of their optimization not considering fairness and the constraints imposed by the platforms;
API solutions have a universality that presents a generalism and perception of the service provider, who often offers better solutions applied to their language or geography;
Managed total AI services raise two questions that highlight their main problems: representative data for whom? fairness how?

For the authors, the solutions/mitigation to the problems raised are presented in three dimensions: from the provider perspective, the user perspective, and the legal perspective. From the provider perspective, they consider the need for providers to adapt to the diversity of user needs, transparency to allow users to review and understand the decisions made, and oversight in the use of their services. From the user perspective, they address responsibility in the contracting and use of AI tools, and an understanding of their needs and challenges in using AI providers. Finally, from the legal perspective, legal support is required to define profiles and responsibilities of both users and providers regarding the results of using the AI service; formalization through user guides and regulation of this type of service; and a form of oversight of this proposed regulation that monitors, guarantees, and enforces this regulation.

Building on the work of \textbf{Yu et al.}, \cite{yu2023antifake}, a solution to mitigate audio deepfake problems that occur from samples published on the internet. These deepfakes can lead to financial problems, unauthorized access, and public image issues.

The computing power GenAI brings to the end user has also led to problems with the misuse of this tool, one of which is audio deepfakes. The availability of audio samples, especially of celebrities, allows these samples to be used to generate deepfakes with their voices. This type of deepfake can be used to commit various types of security breaches.

Some of these breaches are related to voice-based authentication systems, which are fooled by deepfakes, allowing unauthorized user access. Another problematic situation is the use of deepfakes in audio calls, where the attacker can use a relative's voice to request a financial transaction.

As a solution, the authors present a way to pre-process the samples published on the web. This pre-processing algorithm prevents the generation of deepfake audio that is too close to reality. This security comes from inserting information into the audio that makes it difficult to reproduce the voice characteristics of the people being impersonated.

For \textbf{Hutiri et al.} \cite{hutiri2024not}, the use of \textit{deepfakes} for attacks, for example through phone calls, can generate various types of problems for society, ranging from a perception of insecurity on the part of citizens to the unavailability of some resources used to serve citizens (i.e., ambulances). In this scenario, the authors proposed a taxonomy to characterize and identify different types of damage caused by this type of deepfake use.

The authors developed a classification based on three dimensions: type of damage, specific damage, and motivation for the damage. The type of damage is composed of 13 categories, specific damage is composed of 31 different types of damage, and the motivation for the damage is classified into 10 different categories. They also present four different ways in which entities are affected by voice generation (subject of generation, interacting with one, suffering due to a generation, being excluded from a generation) and two ways in which entities can be responsible for attacks (generation and dissemination).

For the authors, the solution to this type of situation involves creating a current taxonomy to understand better the impact of using voice generation. Additionally, they emphasize the importance of classifying output types and assessing whether they can cause harm through exposure to the owners of the voices used in the generation, as well as other potential harmful uses. Furthermore, there is a need for legal support to ensure accountability and regulation of this type of audio generation service, combined with civil society oversight.

In \textbf{Capasso et al.} \cite{capasso2024comprehensive}, a survey overviews the main threats, techniques, prevention methods, and identification methods for computer-generated images. The authors offer a perspective on image forensics and discuss the issues affecting this area of security.

According to the authors, some of the main problems raised are the generation of images of minors and children to promote child prostitution. They also address the challenges of using this type of fake for political purposes, biometric image falsification, system authentication, school activities, and even scientific publications.

Different solutions are presented, but specific to each application scenario. The use of a single solution that continuously identifies the counterfeit is not everyday, but has been pursued by research that uses convolutional neural networks for this purpose.

In their work, \textbf{Liu et al.} \cite{liu2024provcam}, the authors present the problems of disseminated video deepfakes. With advances in GenAI models, videos are becoming increasingly realistic, and detection techniques based on anomalies or physical characteristics of images may soon become ineffective. It is a never-ending tug-of-war.

With video deepfakes, people can easily be impersonated. An example presented by the authors was the situation in Ukraine and Russia, where the president of the former had a fake video go viral. This type of situation can harm political careers and ordinary citizens and have several legal consequences.

The proposed solution was to use secure hardware as a camera module that would apply encryption techniques to ensure the image's origin. The proposal was validated using a Field Programmable Gate Array (FPGA) to simulate the secure component and ensure a valid image sequence in the video.

The work by \textbf{Korobenko et al.} \cite{korobenko2024towards} provides an overview of the ethics of AI use and how countries are approaching the issue. The authors cite the 11 principles proposed by \cite{jobin2019global}: transparency, justice and fairness, non-maleficent, responsibility, privacy, beneficence, freedom and autonomy, trust, dignity, sustainability and solidarity. However, for their work, the authors group some concepts into a set of only five principles: transparency and explainability, privacy and security, fairness and justice, responsibility and accountability, and freedom and autonomy.

This work is categorized into four dimensions: data, technology, people, and processes. The problems are classified according to these dimensions, as are the proposed solutions. For the data dimension, we have concerns about data classification, management, and protection. For technology, the focus is on resource protection and application security. The people dimension focuses on authentication, authorization, and stakeholder engagement. Finally, for processes, the spotlight is on compliance, incident detection, and management.

The authors present a set of five recommendations aimed at ensuring privacy and limiting access to organizational or personal information. These recommendations address everything from data privacy to the governance of processes that use AI.

Continuing with \textbf{Ristovska}, in her work \cite{ristovska2023ways}, the author presents the implications of GenAI for legal aspects in the evaluation of video-based evidence. She also presents some cases of using unsecured video evidence for police decision-making and as confession evidence during interrogations. According to the author, many challenges are associated with this type of evidence in the era of digital media synthesis.

The main problems are the lack of safeguards for situations where there is a possibility of a deepfake in video evidence. This type of safeguard does not yet exist in countries with more advanced development in GenAI, such as the USA, and it may take time to be implemented in countries with fewer resources. This makes it clear that the legal system in most countries is not prepared to mitigate this type of situation, and that one possible outcome is the questioning of decisions made based on this type of digital video technology.

As a solution, the author argues that the real need for using video as evidence should be evaluated, given the current difficulty in authenticating it. She also comments on the need to develop safeguards and that the legal system must be better prepared for this type of situation. Equally important, the author comments on the importance of adopting new technologies to address the challenges presented in her work, challenges associated with the use of video as evidence and the proof of its authenticity.

Following the analysis, the work by \textbf{Han et al.} \cite{han2024uncovering} does not directly focus on problems in understanding the characteristics perceived by people in audio. These characteristics should be used to advance science, especially for the detection of computer-generated voices. This work conducted a study with both sighted and blind people. For the latter, audio communication and the nuances of the voice provide all the feedback that complements the transmitted message.

This work focuses on the characteristics of voices and improving their recognition. Some characteristics are fundamental for identifying the authenticity of an audio, and this work presents six of these characteristics: accent, vocal inflection, signs of liveness, the presence of human error, and acoustic properties. This type of identification complexity can pose problems for blind people who can only rely on voice aspects to identify individuals and emotions.

Some of the solutions address the issue of digital watermarks, which should also be perceptible to humans. Additionally, the evolution of feature classifiers, such as emotion classifiers, and the use of ensemble classifiers to address more situations are also considered. Equally important, the development of technologies that guarantee the identification of the audio generation source and the use of perceptible watermarks can help mitigate this type of problem.

The study \cite{punnappurath2024advocating} by \textbf{Punnappurath et al.} presents a proposal for identifying authenticity in images at the pixel level. Since GenAI can sometimes hallucinate during certain moments of image generation, it is possible to identify generated/altered parts of the image and the original parts. Furthermore, a pixel-level authenticity mask that contains authentication metadata is suggested for this purpose. Finally, post-processing procedures on images, especially on cell phones, can significantly alter their content.

Failure to identify this type of image generation or alteration can lead to a situation where a deepfake is not identified. This type of problem has occurred repeatedly through the publication of AI-edited or transformed images on social media. Another possible threat is the generation or alteration of images that could violate rights or denigrate the image of another citizen. Alternatively, as in the examples shown, due to hallucinations during image reconstruction, AI can alter the content of an image, such as text or numbers.

As a solution, the work proposes implementing pixel-level validation at the time of image acquisition to prevent hallucinations from embedded AI modules. It also proposes adding metadata describing such changes and a spatial mask to identify pixels affected by AI module corrections.

In their work \cite{bird2024cifake}, \textbf{Bird et al.} present a dataset with a total of 120,000 authentic and generated images for use in synthetic data authenticity tools, specifically photographs. The work also proposes a method for identifying AI-generated images and the use of XAI for complex processes involving AI-generated images.

This work addresses the problem of distinguishing between AI-generated and authentic images. This problem has become more significant with the evolution of image generation models and the increasing quality of the images generated by such models. Even for experts, it has become challenging to identify what is real and what is not, leading to fundamental ethical, social, professional, and legal problems in the use of technology.

As a solution, we use the proposed database, along with XAI and a fake identification method that uses the CIFAKE dataset. In their identification solution, the authors use a Grad-CAM-based classification technique.

%
For \textbf{Shoaib et al.} \cite{shoaib2023deepfakes}, preventing the use of deepfake generation is important to avoid cyberscamming and cyberbullying. In their work, the authors propose a framework that integrates policies, detection, and cross-platform collaboration to mitigate the harm caused by GenAI content.

Their problems are linked to cyber-scamming and cyberbullying. Scamming enables various types of attacks that can cause financial, moral, or reputational damage. Cyberbullying can lead people, especially adolescents, to suffer harassment and damage to their image at a time in their lives when they are most vulnerable. In general, the paper lists the social harms as: effects on democracy and public opinion, impacts on privacy and personal security, consequences for the media and journalism, erosion of public trust, and ethical and legal dilemmas.

As a solution, the authors present some technical defense mechanisms, cross-platform strategies, and ethical aspects—all integrated into their framework. As for the technical mechanisms, we can mention detection algorithms, AI-driven authentication methods, and machine learning-based authentication techniques. Their framework proposes to create a form of collaboration between different entities and technologies to avoid this type of problem.

%
According to \textbf{Bianco et al.} \cite{bianco2023identifying}, identifying artworks has become a challenge following the advent of digital media synthesis with Generative Adversarial Networks (GAN) and Diffusion Models. The authors comment that with this type of technology, it is possible to imitate artists' techniques with a high level of fidelity, making it very difficult to distinguish between authentic and AI-generated artworks.

One problem is the issue of evaluation, pricing, and authenticity of works of art. Since AI-generated works are highly faithful to those produced by humans, this has become a problem for evaluators of this type of artifact.

As a solution, the creation of a dataset with images of artworks and the use of XAI and Grad-CAM to classify the works to be analyzed were proposed. It also proposes an ensemble approach for future work to improve results by combining multiple models.

\section{Conclusion and Future Work}
\label{sec_conclusion}

This work satisfactorily addressed the state of the art through a comprehensive scoping review of this research area. This review presented a portfolio that encompasses various aspects of authentication and authenticity that are of great importance when using Generative AI. It also provided a discussion of these aspects, highlighting the most relevant work, its challenges, the targeted attack surfaces, the main threats, solutions, and gaps associated with the topic. The final section concludes with a more detailed analysis of the existing work in the portfolio.

The analysis revealed many problems associated primarily with fraud and fakes in their use. However, beyond algorithms, solutions are presented that encompass public policies, laws, education, and self-awareness. This demonstrates that the solution must integrate various technological, legal, and social aspects, requiring an approach that does not yet exist.

As for future work, we emphasize the importance of detailing the challenges within each sub-area of GenAI applications. More specifically, it is important to detail text generation for different media (i.e., social media), literary texts, audio for social media, audio for phone calls, fully computer-generated images, GenAI-enhanced images, photos, devices with AI-enhanced photos, videos, and others. Because this is a significant new area of activity, a large body of new work is needed to investigate further the threats and solutions that affect each type of affected media.
\bibliographystyle{ACM-Reference-Format}
\bibliography{sample-base}

\appendix

\section{Portfolio Data Extraction}
\label{sc:data_extraction}

\begin{table}[h]
    \centering
    \begin{tabular}{lll}
    \toprule
    \textbf{\#} & \textbf{Field} & \textbf{Decription} \\
    \midrule
        F01 & Identifier & sequential number to identify the field to be studied\\
        F02 & Authors & names of the authors involved in the publication\\
        F03 & Title & the original name of each selected article\\
        F04 & Problem & the problem addressed in the work, from a broader perspective\\
        F05 & Object & the specific object of study of this work, from a more focused perspective\\
        F06 & Context & the context of the study or application of the solution\\
        F07 & Study & the type of study conducted\\
        F08 & Area & the area of science in which the study was conducted\\
        F09 & Theme & the theme of the work\\
        F10 & Approach & which approach, technique, or algorithm was used to solve the problem\\
        F11 & Database & did you use any databases to evaluate the solution? Which ones?\\
        F12 & Participants & how many participants were involved in the study?\\
    \bottomrule
    \end{tabular}
    \caption{Set of fields used to extract data from articles.}
    \label{tab:field_description}
\end{table}

\section{Final Portfolio}
\label{sc:final_portfolio}

\begin{table}
\centering
\begin{tabularx}{.9\textwidth}{lXX}
\toprule
\textbf{\#} & \textbf{Author} & \textbf{Article} \\
\midrule
    1  & Datta, Pratim, Whitmore, Mark, Nwankpa, Joseph K.                                                 & A Perfect Storm: Social Media News, Psychological Biases, and AI                                                     \\
    2  & Lewicki, Kornel, Lee, Michelle Seng Ah, Cobbe, Jennifer, Singh, Jatinder                          & Out of Context: Investigating the Bias and Fairness Concerns of “Artificial Intelligence as a Service”               \\
    3  & Yu, Zhiyuan, Zhai, Shixuan, Zhang, Ning                                                           & AntiFake: Using Adversarial Audio to Prevent Unauthorized Speech Synthesis                                           \\
    4  & Hutiri, Wiebke, Papakyriakopoulos, Orestis, Xiang, Alice                                          & Not My Voice! A Taxonomy of Ethical and Safety Harms of Speech Generators                                            \\
    5  & Capasso, Paola, Cattaneo, Giuseppe, De Marsico, Maria                                             & A Comprehensive Survey on Methods for Image Integrity                                                                \\
    6  & Liu, Yuxin (Myles), Yao, Zhihao, Chen, Mingyi, Amiri Sani, Ardalan, Agarwal, Sharad, Tsudik, Gene & ProvCam: A Camera Module with Self-Contained TCB for Producing Verifiable Videos                                     \\
    7  & Korobenko, Daria, Nikiforova, Anastasija, Sharma, Rajesh                                          & Towards a Privacy and Security-Aware Framework for Ethical AI: Guiding the Development and Assessment of AI Systems  \\
    8  & Ristovska, Sandra                                                                                 & Ways of seeing: The power and limitation of video evidence across law and policy                                     \\
    9  & Han, Chaeeun, Mitra, Prasenjit, Billah, Syed Masum                                                & Uncovering Human Traits in Determining Real and Spoofed Audio: Insights from Blind and Sighted Individuals           \\
    10 & Punnappurath, Abhijith, Zhao, Luxi, Abdelhamed, Abdelrahman, Brown, Michael S.                    & Advocating Pixel-Level Authentication of Camera-Captured Images                                                      \\
    11 & Bird, Jordan J., Lotfi, Ahmad                                                                     & CIFAKE: Image Classification and Explainable Identification of AI-Generated Synthetic Images                         \\
    12 & Bianco, Tommaso, Castellano, Giovanna, Scaringi, Raffaele, Vessio, Gennaro                        & Identifying AI-Generated Art with Deep Learning                                                                      \\
    13 &  Shoaib, Mohamed R., Wang, Zefan, Ahvanooey, Milad Taleby, Zhao, Jun                               & Deepfakes, Misinformation, and Disinformation in the Era of Frontier AI, Generative AI, and Large AI Models  \\
\bottomrule
\end{tabularx}
\caption{Final portfolio - set of articles selected to compose the final portfolio of this work}
    \label{tab:final_portfolio}
\end{table}

\section{Acknowledgments}

The authors sincerely thank the Federal University of Santa Catarina (UFSC). This study was partially funded by the Fundação de Amparo à Pesquisa e Inovação do Estado de Santa Catarina (FAPESC), Edital 20/2024.

\end{document}